\documentclass[prd, amsfonts, twocolumn, nofootinbib, showpacs]{revtex4}
\usepackage{graphicx, epsfig}
\usepackage{color}
\usepackage{amsmath}
\usepackage{amssymb}
\newcommand{\be}{\begin{equation}}
\newcommand{\ee}{\end{equation}}
\newcommand{\bea}{\begin{eqnarray}}
\newcommand{\eea}{\end{eqnarray}}

\newcommand{\gapp}{\mathrel{\raise.3ex\hbox{$>$}\mkern-14mu
\lower0.6ex\hbox{$\sim$}}}
\newcommand{\lapp}{\mathrel{\raise.3ex\hbox{$<$}\mkern-14mu
\lower0.6ex\hbox{$\sim$}}}
\def\bbox{{\,\lower0.9pt\vbox{\hrule \hbox{\vrule height 0.2 cm
\hskip 0.2 cm \vrule  height 0.2 cm}\hrule}\,}}

\begin{document}
\title{Hydrogen atom wave function and eigen energy in the Rindler space }
\author{De-Chang Dai}
\affiliation{ Institute of Natural Sciences, Shanghai Key Lab for Particle Physics and Cosmology, \\
and Center for Astrophysics and Astronomy, Department of Physics and Astronomy,\\
Shanghai Jiao Tong University, Shanghai 200240, China}


\begin{abstract}
\widetext
We study the hydrogen atom eigenstate energy and wave function in the Rindler space. The probability distribution is  tilted because the electric field of the nucleus is no longer spherically symmetric. The hydrogen atom therefore cannot be treated exactly in the same way as what it is in an inertial frame. We also find that if the external force accelerates only the nucleus and then the nucleus accelerates its surrounding electrons through electromagnetic force, the electrons can tunnel through the local energy gap and split the hydrogen atom into an  ion. This is similar to what one expects from the Stark effect. However, the critical acceleration is about $3\times 10^{22} m/s^2$. It is well beyond the gravitational acceleration on a regular star surface.
\end{abstract}


\maketitle

\section{Introduction}

Rindler spacetime is described by a coordinate system in which every observer is uniformly accelerated with respect to the Minkowski space (or an inertial frame) \cite{Rindler(1966)}. Every observer undergoes hyperbolic motion in the flat spacetime. Uniform acceleration in Minkowski spacetime plays an important role in understanding the motion in the presence of the horizons and related phenomena. For example, a uniformly accelerating observer observes thermal radiation \cite{Unruh(1976)}. Since acceleration is considered equivalent to gravity, it has been intensively studied in order to understand the similar phenomena, like Hawking radiation\cite{Hawking1,Hawking2}.

One of the technique to extract localized spatio-temporal quantum information is through Unruh-DeWitt Detector\cite{Unruh:1976db,DeWitt}. The idea is to construct a detector which interacts with the quantum field. The excitation of the detector is interpreted as the absorption of a particle. It is generally assumed that the detector is not affected by the magnitude of the acceleration. This is very hard to be true in a real detection. Any detector with finite size,  eg. atom, will be affected by the acceleration and changes its structure. For example, the Van der Waals force between accelerated molecules are not the same as which between molecules without acceleration\cite{Noto:2013ona}. An accelerated charged particle also encounters a self-force which is caused by its own electric field\cite{Fermi,Frolov,Gralla:2009md,Pinto2006}. This makes it even harder  to study an accelerated atom's radiation shift\cite{Passante(1998),Belyanin}.

Hydrogen atom in a curved spacetime have been studied for the case of  Schwarzschild, De-Sitter and some general metric\cite{Parker(1980),Parker:1980kw,Parker:1982nk,Zhao:2007xj,Zhao:2007as,Moradi(2010)}. These studies focused on a hydrogen atom in a free fall or Riemann normal coordinates. In contrast, we here focus on the acceleration effects on hydrogen. Though acceleration can be locally considered equivalent to gravity, acceleration from internal and external sources can have different effects. To the best of our knowledge, this case has not been previously studied in the literature.

In one of the cases that we consider, an electron within the hydrogen atom is accelerated only by the proton, while in the other case the electron is accelerated by both the proton and an external force. These two cases have different acceleration sources and therefore they should have different effect on the hydrogen atom structure. To simplify the study, We neglect the electron's spin and describe the electron with a charged scalar field. The first step is to find the hydrogen's Schrodinger equation in the Rindler spacetime. A free particle's Schrodinger equation has been studied for a while and it has been found that there could be a new type of quanta in the Rindler spacetime\cite{De:2015dja,Mitra:2016fzi}. We extend the study to a scalar field particle which is bounded by the atom's nucleus. This nucleus is accelerated uniformly by an external force and its electric field is static in the Rindler spacetime. The solution has been found by Whittaker\cite{Whittaker} and it is no longer spherically symmetric (see fig. \ref{contour}). The electron is bound by the nucleus. The electron's acceleration can be provided by either the nuclei or an external force. If the electron is accelerated by the nucleus, then the electron's probability distribution is pulled downward by the gravity in the Rindler spacetime. This is exactly what one expects. And if the gravitational acceleration is big enough, the atom can be split and a bound hydrogen can not be formed. The critical acceleration is  $a\approx 3\times 10^{22} m/s^2$. This is unlikely to happen in a regular star's surface.

The other case is that the electron is also accelerated by an external force. This force cancels out the gravity in the Rindler spacetime. The nucleus' electric field can attract electrons without gravitational resistance. It turns out that the probability distribution is tilted upward (opposite to gravity). This result implies that if a detector is made of atoms, its responding energy will not the same as it is in an inertial frame.  The concept of an Unruh-DeWitt detector is invented to study how a detector responds to an accelerated source. The detector response will depend on how the external force interacts with the detector. For example if the detector is accelerated as a whole by an external force, its wavefunction must be distorted in a way similar to the second case in this paper. In the following, we first introduce a static charged particle's electric field in the Rindler spacetime. The Schrodinger equation is obtained by the approximation of Klein-Gordon field. The energy eigenvalues and eigen states are obtained by the perturbation theory.

\section{A charged particle's electric field in the Rindler space}

In Rindler spacetime all the observers move with a uniform acceleration with respect to an inertial observed. It can be written down in a static form as

\begin{equation}
\label{matric}
ds^2=(1+az)^2 dt^2 -(dx^2+dy^2+dz^2)
\end{equation}

Here we take $c=\hbar=1$. Here $a$ is the magnitude of the acceleration and it is directed to the $+z$-direction. A free particle experiences a uniform force toward $-z$-direction. This force is similar to gravitational acceleration and mimics a uniform gravitational field.

Now we consider a charged particle, with charge $q$, which is accelerated under a uniform acceleration. It stays at a fixed point in the Rindler space. We place the particle at $(x=0,y=0,z=0)$ for convenience. This charged particle generates an electromagnetic vector potential $(A_t,A_x,A_y,A_z)$. In the Rindler space the vector field is \cite{Pinto2006,Whittaker}

\begin{eqnarray}
\label{vector1}
&&A_t(r)=\frac{q}{r}\frac{1+az+\frac{a^2}{2}r^2}{\sqrt{1+az+\frac{a^2}{4}r^2}}\\
\label{vector2}
&&A_x=A_y=A_z=0
\end{eqnarray}

 Here, $q$ is the proton's charge. 

\begin{figure}[ht!]
   \centering
\includegraphics[width=6cm]{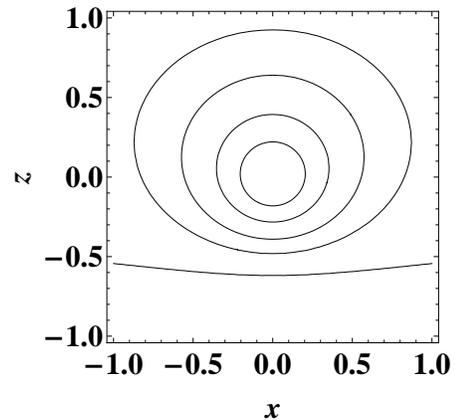}
\caption{This is the equivalent potential contour of a charged particle's vector field's time component or electric field, $A_t$. The acceleration is chosen to be $a=1$. x-axis and z-axis' units are $\frac{1}{a}$. The contour in the lower part (negative $z$) is denser than the contour field in up part. The field in the lower part increases much quicker than that in the upper part.}
\label{contour}
\end{figure}

One finds the vector field, $A_t$, in the lower part change much quicker than that in the upper part(fig \ref{contour}).

\section{Hydrogen wave function in the Rindler space time}

A hydrogen atom is made of a proton and an electron. The proton is much heavier than the electron. Therefore the proton can be treated as a static charged particle. In the Rindler space this proton provides an electric potential $A_t$. The electron is affected by the field and moves around. Similar case has been studied by Lenz et al.\cite{Lenz}. To simplify the calculation, we neglect the electron's spin and assume it is described by a charged scalar field. The electron's action is

\begin{eqnarray}
S_1&=&\int \Big((D_\mu-ieA_\mu) \phi^* (D^\mu+ieA^\mu)\phi \nonumber\\
&&-m^2 \phi^*\phi -\frac{1}{4}F^{\mu\nu}F_{\mu \nu}\Big) \sqrt{-g}d^4x
\end{eqnarray}

The electron's equation of motion is

\begin{equation}
(D_\mu+ieA_\mu)  (D^\mu+ieA^\mu)\phi +m^2 \phi=0
\end{equation}

One applies the geodesic metric (equation \ref{matric}) into the above equation,

\begin{equation}
\frac{1}{(1+az)^2}(\partial_t+ieA_t)^2\phi-\nabla^2\phi -\frac{a}{1+az}\partial_z\phi  +m^2 \phi=0
\end{equation}

Here $\nabla^2=\partial_x^2+\partial_y^2+\partial_z^2$

We keep only zeroth and first order terms in $a$ and rewrite the equation in the non-relativistic approximation. In this case the equation becomes the Schrodinger equation with a perturbation term.

\begin{eqnarray}
i\partial_t\phi&=& H_0\phi+aH_1\phi+O(a^2)\\
H_0&=&-\frac{\nabla^2}{2m}   +m +\frac{qe}{r}\\
\label{pert1}
H_1&=&-z\frac{\nabla^2}{2m}-\frac{\partial_z}{2m}+mz +\frac{z}{2}\frac{qe}{r}
\end{eqnarray}

Here we keep the rest mass energy. The third term, $mz$, in equation \ref{pert1} is corresponding to the classical gravitational potential. The electron is pulled down by the gravity and the proton makes the electron accelerate together with it through  it's electromagnetic force. In other words, the electron is accelerated by the proton and the proton is accelerated by an external force. This is different from the case that in which an electron is also accelerated by some external force. We will discuss this in the next section.
The other terms in equation \ref{pert1} are much smaller than the third term and can be neglected. Once these terms are neglected, the perturbation potential is similar to the Stark effect.
\begin{equation}
V\approx \frac{qe}{r}+amz+m
\end{equation}

\begin{figure}[ht!]
   \centering
\includegraphics[width=6cm]{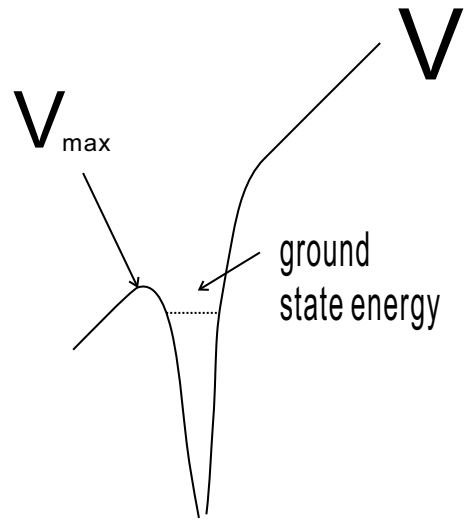}
\caption{The black curve is the potential of an accelerated electron.  The horizonal direction is the z-axis. There is a local minimum at the proton's location. The electron can be captured and form a hydrogen atom, if the hydrogen's ground state energy is lower than the local maximum $V_{max}$, the electron can also tunnel through the potential barrier and the hydrogen atom becomes a single proton. }
\label{potential}
\end{figure}

The potential has two local minima. One is at the proton's location and the other one is at $-\infty$ (see fig. \ref{potential}). It is known that an electron can tunnel through the local potential barrier near the proton and becomes free\cite{griffiths}, and the hydrogen atom becomes an ion. This process is called field-ionization\cite{Yamabe,Smirony,Banks,Bailey,Littman}.  The process is very complicated and involves tunneling. However, if the potential barrier, $V_{max}$, is higher than the atom's ground state, the electron will definitely escape. Griffiths gives a simple estimation of tunneling time\cite{griffiths}.

\begin{eqnarray}
\tau&=&\frac{8mw^2}{\pi}\exp(2\gamma)\\
\gamma&=&\frac{2\sqrt{2m}}{3ma}V_0^{\frac{3}{2}}
\end{eqnarray}

$V_0$ is the ground state energy. It is about $13.6$eV. $2w$ is the width of potential well and is about the hydrogen atom's radius. $\exp(2\gamma)$ is the most important suppression factor. The tunneling rate  becomes quick enough to separate the atom if $\gamma\approx 1$. In this case the acceleration must satisfy

\begin{equation}
a\gtrapprox \frac{2\sqrt{2}}{3\sqrt{m}}V_0^{\frac{3}{2}}
\end{equation}

For the hydrogen's ground state, the critical acceleration is about $\approx 3\times 10^{22} m/s^2$ (on a neutron star's surface it is about $10^{12}m/s^2$). It is therefre unlikely to happen near the regular star surface. Therefore it is quite safe to say that a hydrogen is not ionized by the gravitational force along, unless it is very close to a black hole. However, we must point out that the atoms at the star's surface are generally supported by atoms' electrons interacting with the other particles, and then the electrons interact with their central nuclei. This is opposite to the case we were discussing here. A nucleus is much heavier than an electron, so the critical acceleration must be less than what this estimate provides.

We will now apply the perturbation theory to find the hydrogen eigenenergy and eigenstate. The flat space hydrogen eigenenergy, $H_0$, is

\begin{equation}
E_n^0=\frac{-1}{2ma_0^2n^2}+m
\end{equation}

Its eigenfunctions are

\begin{eqnarray}
&&\phi^0_{n,l,m}(r,\theta,\phi)=R_{n,l}Y^m_l(\theta,\phi)\\
&&R_{n,l}\sim\exp(-\frac{r}{na_0})\Big(\frac{r}{na_0}\Big)^l L^{2l+1}_{n-l-1}(\frac{2r}{na_0})
\end{eqnarray}

$L^{2l+1}_{n-l-1}(x)$ is the associated Laguerre polynomial. $Y^m_l(\theta,\phi)$ is the spherical harmonic function of degree $l$ and order $m$. $a_0$ is the Bohr radius, $a_0=\frac{1}{me^2}=5.2917721067\times 10^{-11}$m. The ground and first excited states are

\begin{eqnarray}
\phi^0_{1,0,0}&=&\Big(\frac{1}{\pi a_0^3}\Big)^{\frac{1}{2}}\exp(-\frac{r}{a_0})\\
\phi^0_{2,0,0}&=&\Big(\frac{1}{32\pi a_0^3}\Big)^{\frac{1}{2}}\Big(2-\frac{r}{a_0}\Big)\exp(-\frac{r}{2a_0})\\
\phi^0_{2,1,0}&=&\Big(\frac{1}{32\pi a_0^3}\Big)^{\frac{1}{2}}\frac{r}{a_0}\exp(-\frac{r}{2a_0})\cos\theta\\
\phi^0_{2,1,\pm 1}&=&\mp\Big(\frac{1}{64\pi a_0^3}\Big)^{\frac{1}{2}}\frac{r}{a_0}\exp(-\frac{r}{2a_0})\sin\theta e^{\pm i\phi}
\end{eqnarray}

From the perturbation theory, the first order correction to the energy is

\begin{eqnarray}
\label{energy1}
&&E_n=E^0_n+a<\phi^0_{n,l,m}|H_1|\phi^0_{n,l,m}>+O(a^2)
\end{eqnarray}

and its first order correction to the eigenfunction is \cite{Shankar}

\begin{equation}
\phi_{n,l,m}=\phi^0_{nlm}+a\sum_{n'\neq n}\frac{<\phi^0_{n',l',m'}|H_1|\phi^0_{n,l,m}>}{E_n^0-E_{n'}^0}\phi^0_{n',l',m'}+O(a^2)
\end{equation}

 One may calculate energy density from the energy momentum tensor after obtaining the wavefunction.

\subsection{Ground state}

The ground state is not a degenerate state. The energy eigen function can be obtained from equation \ref{energy1} directly,

\begin{eqnarray}
\label{energy2}
&&E_1=E^0_1+O(a^2)
\end{eqnarray}

The first order correction is zero. However, the first order wave function correction is not zero. It is

\begin{equation}
\label{series1}
\phi_{1,0,0}=\phi^0_{100}+aa_0\sum_{n=2}^{\infty}b_n\phi^0_{n,1,0}+O(a^2)
\end{equation}

\begin{figure}[ht!]
   \centering
\includegraphics[width=6cm]{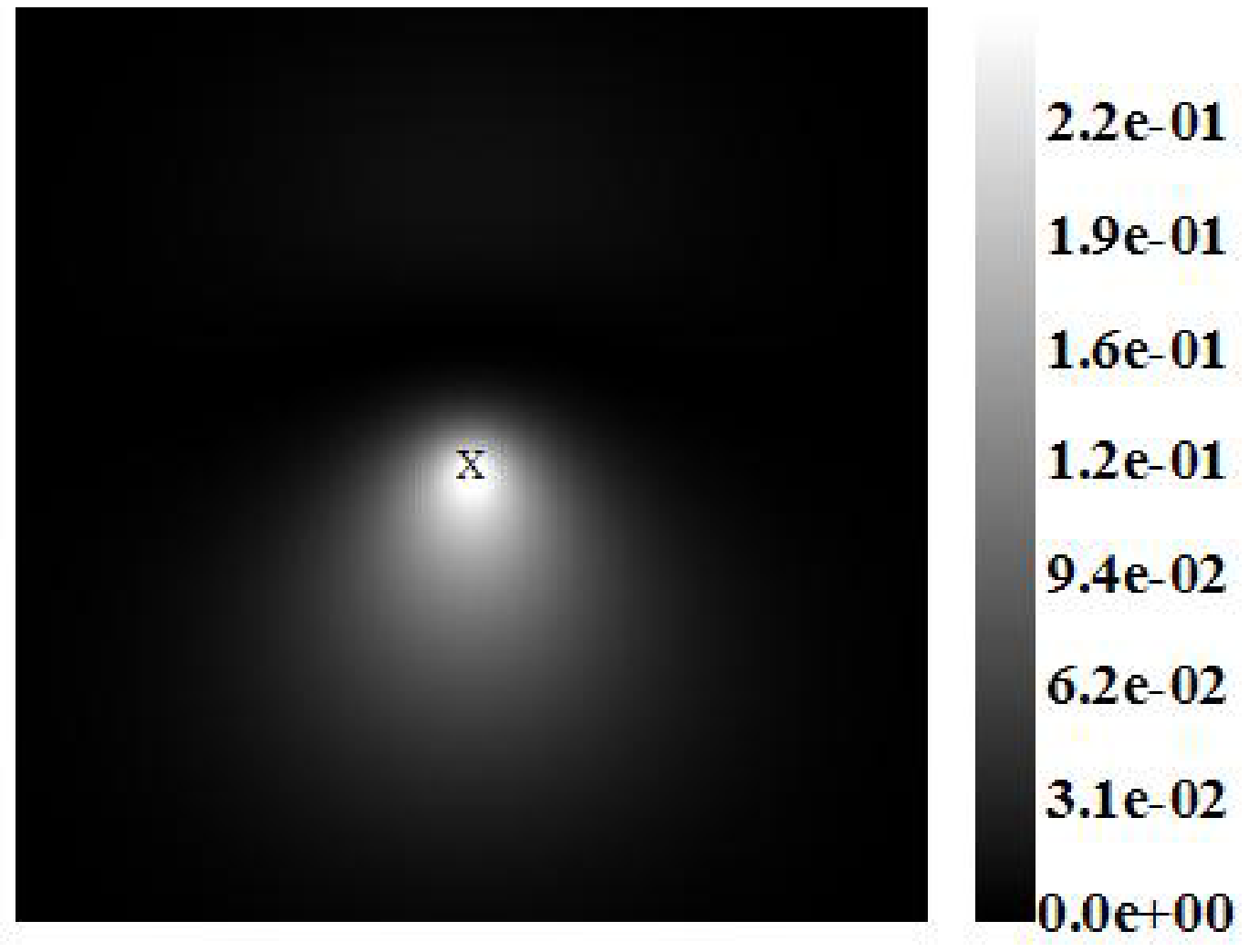}
\caption{This is the ground state probability distribution. The electron overcomes the gravity through the interaction with the central proton. The X in the center marks the location of the proton. The vertical direction is $z$-direction and the horizontal direction is $x$-direction. The full size is $8 a_0\times 8 a_0$. In order to see the distortion, we choose $aa_0=3\times 10^{-7}$.}
\label{ground-noncouple}
\end{figure}

The first 5 coefficients, $b_n$, can be found in table \ref{tab:coe1}. As expected, the wave function is pulled downward by the gravity (see fig. \ref{ground-noncouple}).

\begin{table}
\caption{\label{tab:coe1} The first 5 coefficients of the ground state wave function (equation  \ref{series1}) }
\begin{tabular}{ |c| c| c| c| c|}
 \hline
$b_2$ & $b_3$ & $b_4$ & $b_5$ & $b_6$ \\
 \hline
 $-3.73\times 10^4$ & $-1.26\times 10^4$ &$-7.04\times 10^3$ & $-4.71\times 10^3$ & $-3.46\times 10^3$\\
  \hline
\end{tabular}
\end{table}

\subsection{First excited state}
Unlike the ground state, the first excited state is degenerate and we have to reorganized the wave function bases. We first find the degenerate energy matrix,

\begin{eqnarray}
&&<2ml| H_1|2m'l'>=\nonumber\\
&&\mbox{\bordermatrix{
nml & 200 & 210 & 211&21-1\cr
200 & 0 & -1.53\times 10^6 a_0& 0&0\cr
210 & -1.53\times 10^6 a_0& 0 &0 &0\cr
211 & 0 & 0 &0 &0\cr
21-1 & 0 & 0 &0 &0\cr}}\nonumber\\
\end{eqnarray}

\begin{figure}[ht!]
   \centering
\includegraphics[width=6cm]{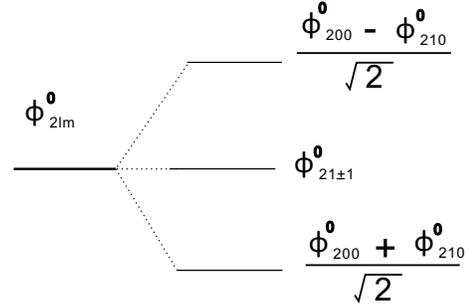}
\caption{The first excited state wave function with respect to the energy split (see equation \ref{sep1}): The electron is accelerated by the proton.}
\label{energy-noncouple}
\end{figure}

\begin{figure}[ht!]
   \centering
\includegraphics[width=6cm]{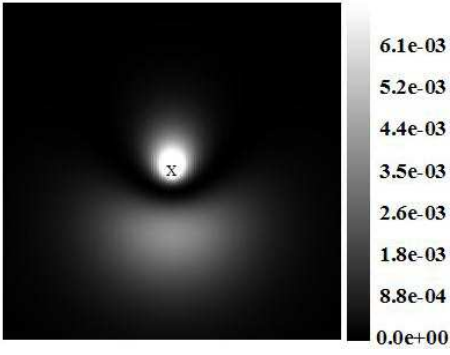}
\caption{This is the first excited state's probability distribution, $\phi^+$. The electron overcomes the gravity through the interaction with the central proton. The X in the center marks the location of the proton. The vertical direction is $z$-direction and the horizontal direction is $x$-direction. The full size is $16 a_0\times 16 a_0$. }
\label{excited-noncouple}
\end{figure}

$\phi^0_{200}$ and $\phi^0_{210}$ are mixing together. So the wave function basis must be reorganized to

\begin{equation}
\phi^\pm=\frac{\phi_{200}^0\pm\phi_{210}^0}{\sqrt{2}}
\end{equation}

and their eigenenergies are

\begin{equation}
\label{sep1}
E_n=E_2^0\mp 1.53\times 10^6aa_0
\end{equation}
The unit is eV. $\phi^+$'s energy is lower than $\phi^-$ so that the state is preferred (see fig. \ref{energy-noncouple}). Fig. \ref{excited-noncouple} shows that $\phi^+$ is pulled upward a little bit near the center, but some of them are pulled down to the much further lower part. This lowers down the system's energy.

\section{The hydrogen wave function with electron pushed by an external force}
We discussed the case where electron is pulled by the proton in the previous section. We now focus on the case where both protons and electrons are also accelerated by an external force. We assume the external force is a vector field $B_\alpha$ (the coupling constant is including in $B_\alpha$). This vector field provides the electron a uniform acceleration so that it can overcome gravitational force. The derivative of the precise form of $B_\alpha$ can be found in the appendix.

The electron's action is written down in the form
\begin{eqnarray}
S_2&=&\int \Big((D_\mu-ieA_\mu-iB_\mu) \phi^* (D^\mu+ieA^\mu+iB^\mu)\phi \nonumber\\
&&-m^2 \phi^*\phi -\frac{1}{4}F^{\mu\nu}F_{\mu \nu} \Big) \sqrt{g}d^4x\\
B_t&=&-maz\\
B_x&=&B_y=B_z=0
\end{eqnarray}
 $B_\alpha$'s kinetic terms are neglected in the equation.  $A_\mu$ is the proton's electromagnetic four vector potential. Its form can be found in equation \ref{vector1} and \ref{vector2}.  $\phi$'s equation of motion is

\begin{equation}
\frac{1}{(1+az)^2}(\partial_t+ieA_t+iB_t)^2\phi-\nabla^2\phi -\frac{a}{1+az}\partial_z\phi  +m^2 \phi=0
\end{equation}

We keep only the zeroth and first order terms in $a$ and rewrite the equation in the non-relativistic approximation. In this case the equation can be written in the Schrodinger form as

\begin{eqnarray}
i\partial_t\phi&=& H_0^p\phi+aH_1^p\phi+O(a^2)\\
H_0^p&=&-\frac{\nabla^2}{2m}   +m +\frac{qe}{r}\\
\label{pert2}
H_1^p&=&-z\frac{\nabla^2}{2m}-\frac{\partial_z}{2m} +\frac{z}{2}\frac{qe}{r}
\end{eqnarray}

If we compare equation \ref{pert2} with equation  \ref{pert1}, we can see that $mz$ term is not in equation \ref{pert2}. It is because $B_\alpha $ cancels out gravity and now the electron is accelerated together with the proton.

 Since $H_0^p$ is exactly the same  as $H^0$, we can obtain the eigenenergy and eigenfunction correction term through

\begin{eqnarray}
\label{energy2}
&&E_n^p=E^0_n+a<\phi^0_{n,l,m}|H_1^p|\phi^0_{n,l,m}>+O(a^2)
\end{eqnarray}

The first order correction in the eigenfunction is \cite{Shankar}

\begin{equation}
\phi_{n,l,m}^p=\phi^0_{nlm}+a\sum_{n'\neq n}\frac{<\phi^0_{n',l',m'}|H_1^p|\phi^0_{n,l,m}>}{E_n^0-E_{n'}^0}\phi^0_{n',l',m'}+O(a^2)
\end{equation}



\subsection{Ground state}

The ground state is not a degenerate state. The eigen energy  can be obtain from equation \ref{energy2} directly,

\begin{eqnarray}
\label{energy3}
&&E_1^p=E^0_1+O(a^2)
\end{eqnarray}

The first order correction is zero again. However, the first order wave function correction is also not zero. It is

\begin{equation}
\label{series2}
\phi_{1,0,0}^p=\phi^0_{100}+aa_0\sum_{n=2}^{\infty}c_n\phi^0_{n,1,0}+O(a^2)
\end{equation}

\begin{figure}[ht!]
   \centering
\includegraphics[width=6cm]{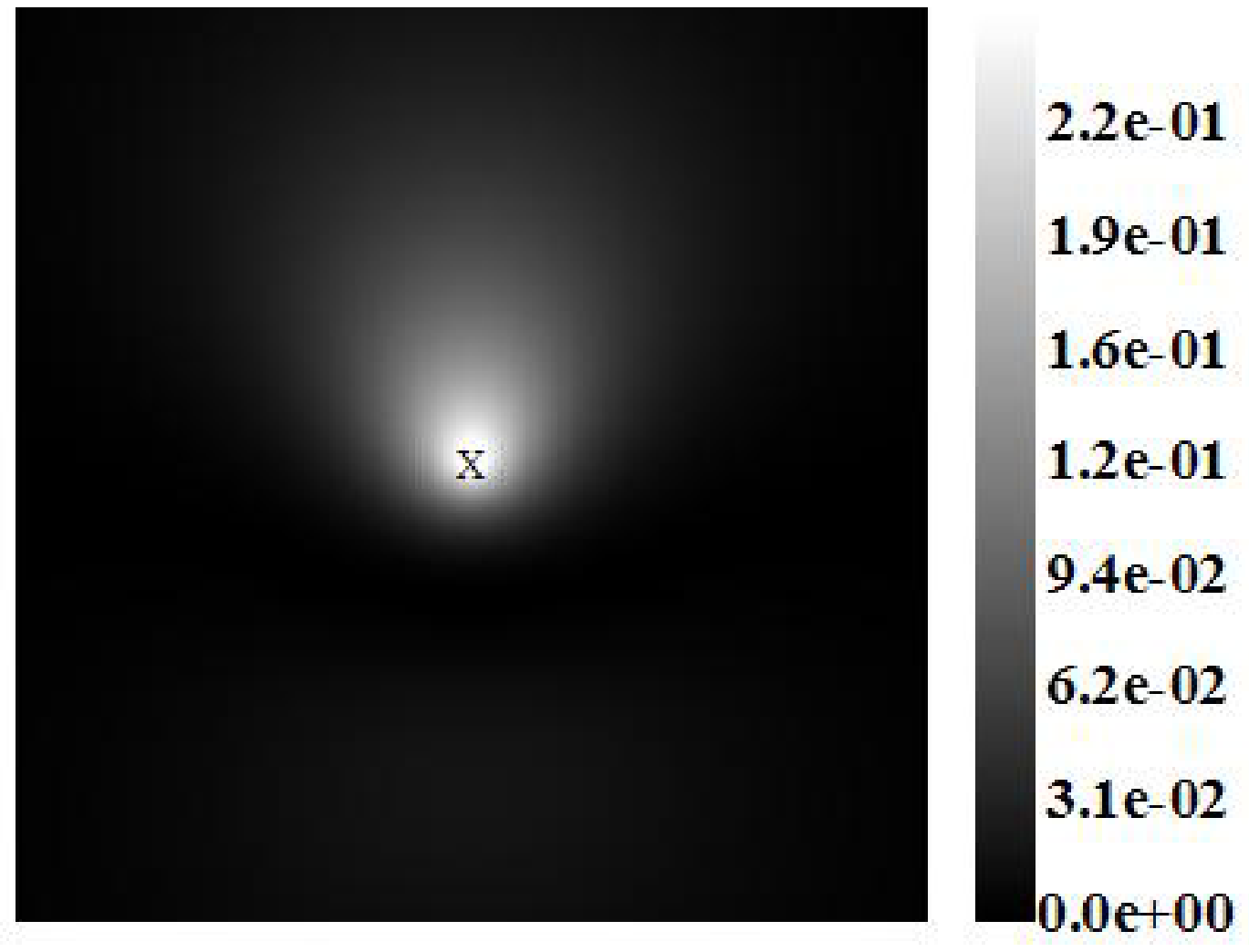}
\caption{This is the ground state probability distribution. The electron overcomes gravity through the interaction with the central proton. The X in the center marks the location of the proton. The vertical direction is $z$-direction and the horizontal direction is $x$-direction. The full size is $8 a_0\times 8 a_0$. In order to see the distortion, we choose $aa_0=2$.}
\label{ground-couple}
\end{figure}

The first 5 coefficients, $c_n$, can be found in table \ref{tab:coe2}. Unlike the previous case, the wave function is pulled upward (see fig. \ref{ground-couple}). This is because that the electric potential at the upper part has much lower potential than the potential at the corresponding lower part( fig.\ref{contour}).

\begin{table}
\caption{\label{tab:coe2} The first 5 coefficients of the ground state wave function (equation  \ref{series2}) }
\begin{tabular}{ |c| c| c| c| c|}
 \hline
$c_2$ & $c_3$ & $c_4$ & $c_5$ & $c_6$ \\
 \hline
 $9.93\times 10^{-1}$ & $3.35\times 10^{-1}$ &$1.87\times 10^{-1}$ & $1.25\times 10^{-1}$ & $9.21\times 10^{-2}$\\
  \hline
\end{tabular}
\end{table}

\subsection{First excited state}
The first excited state is degenerate and one must reorganize the wave function bases. We first find the degenerate energy matrix,

\begin{eqnarray}
&&<2ml| H_1^p|2m'l'>=\nonumber\\
&&\mbox{\bordermatrix{
nml & 200 & 210 & 211&21-1\cr
200 & 0 & 10.2a_0 & 0&0\cr
210 & 10.2a_0 & 0 &0 &0\cr
211 & 0 & 0 &0 &0\cr
21-1 & 0 & 0 &0 &0\cr}}
\end{eqnarray}

$\phi^0_{200}$ and $\phi^0_{210}$ are mixing together. So the wave function basis must be reorganized to

\begin{equation}
\phi^{p\pm}=\frac{\phi_{200}^0\pm\phi_{210}^0}{\sqrt{2}}
\end{equation}

and their eigenenergies are

\begin{equation}
\label{sep2}
E_{2}^{p\pm}=E_2^0\pm 10.2aa_0
\end{equation}

\begin{figure}[ht!]
   \centering
\includegraphics[width=6cm]{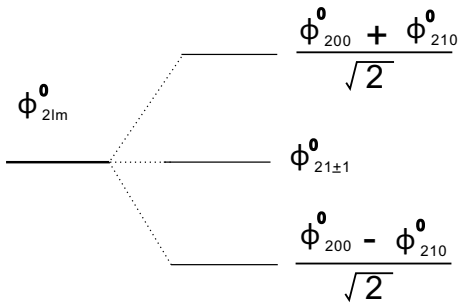}
\caption{The first excited state wave function with respect to the energy split (equation \ref{sep2}): The electron is accelerated by an external vector field.}
\label{energy-couple}
\end{figure}

\begin{figure}[ht!]
   \centering
\includegraphics[width=6cm]{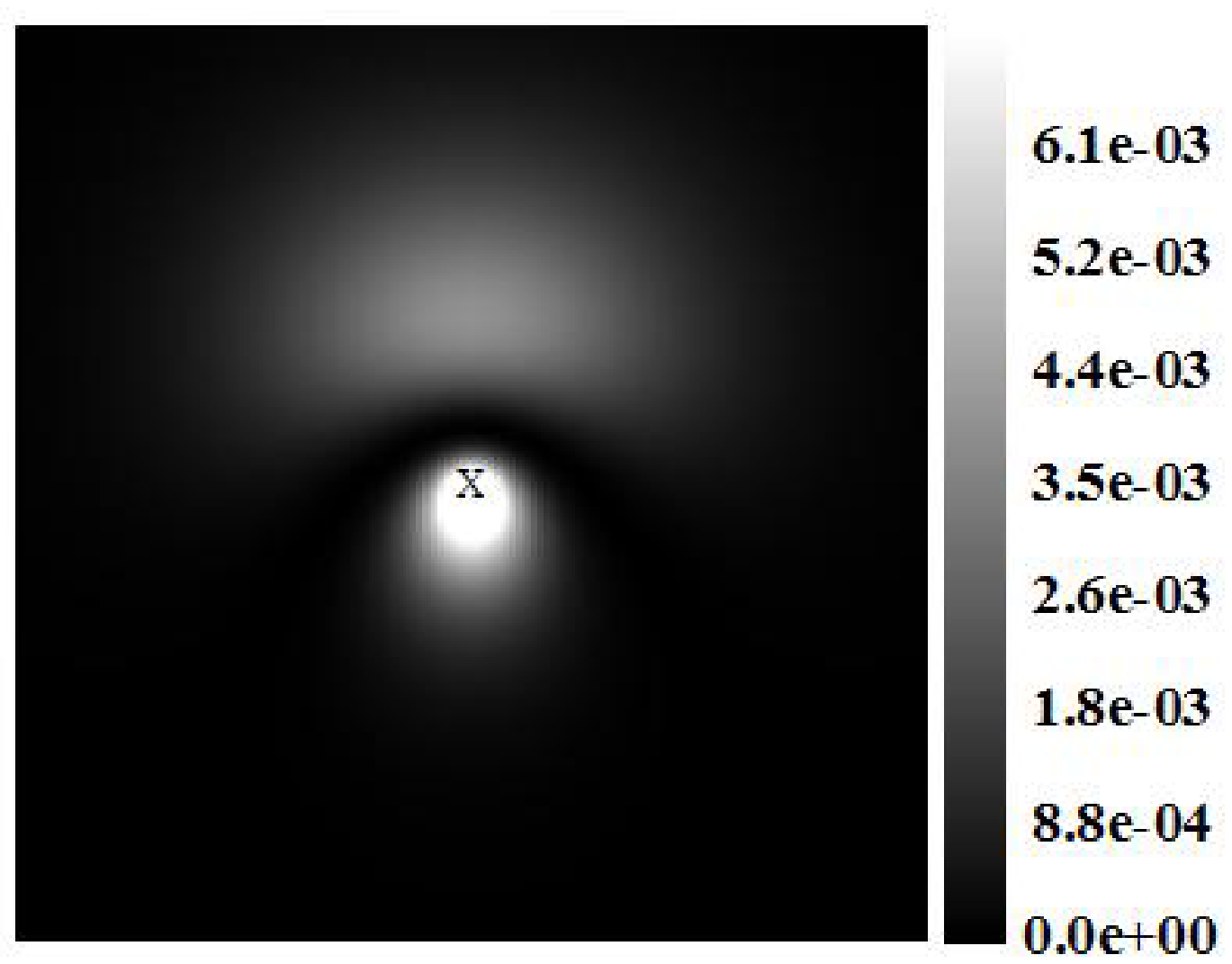}
\caption{This is the first excited state's probability distribution, $\phi^{p-}$. The electron overcomes gravity through an external vector field, $B_\alpha$. The X in the center marks the location of the proton. The vertical direction is $z$-direction and the horizontal direction is $x$-direction. The full size is $16 a_0\times 16 a_0$.}
\label{excited-couple}
\end{figure}

The unit here is an eV. $\phi^{p-}$'s energy is lower than $\phi^{p+}$ so that the state is preferred (fig. \ref{energy-couple}). Fig. \ref{excited-couple} shows that $\phi^{p-}$ is pulled downward a little near the center, but some of them are pulled upward to the much further upper part. This lowers down the system's energy.

\section{Conclusion}

In this paper, we discussed the hydrogen's wave function in the Rindler spacetime. We separate the analysis to two cases. The first case is when the central proton is accelerated by an external force, but the electron is not accelerated by that external force. The electron  interacts only with the proton through the electromagnetic force. In this case the electron is pulled downward by a gravitational potential, $maz$, since the gravitational potential is the dominant perturbation term. This case is very similar to the Stark effect\cite{griffiths}. This effect will not appear in the case of a freely falling object\cite{Parker(1980),Parker:1980kw,Parker:1982nk}. If the acceleration is strong enough, the bound state cannot be formed and a hydrogen is separated into an ion and electron. A simple estimate shows the critical acceleration is $a\approx 3\times 10^{22} m/s^2$. This is much larger than the gravitational acceleration on a regular star surface. Therefore, it is unlikely to happen unless it is very close to a black hole. However, our estimate is based on the central proton accelerating its surrounding electron. In general case, the atoms at star's surface are supported by atoms' electrons interacting with the other material on the star surface, and then the electrons interact with its central nuclei. The nuclei are therefore supported by their electrons. Since a nucleus is much heavier than an electron, the critical acceleration must be less than what this estimate provides. We then calculate the eigen wave function and eigenenergy via the perturbation theory. We find that the ground state's energy to the first order in perturbation does not change, but its probability function is pulled downward. It is exactly what one expects in a gravitational field. The first excited state is split to 3 different energy  levels. The lowest energy state's probability function is also titled to the gravitational force's direction, although  the probability seems tilted upward near its center.

The second case is when both proton and electron are accelerated by an external force. In this case tunneling will not happen and hydrogen atom can be formed. The shift of the eigenfunction and the eigenenergy is different from the first case. The ground state energy does not change in the first order, but the probability is pulled upward. This is because the potential at the upper side is lower than that at the lower part. The electron will prefer to stay at the upper part to lower its energy. The first excited state is split into three energy states. The lowest energy state's probability  distribution is tilted upward.

  An Unruh-DeWitt detector describes how vacuum responds to an accelerated field. If the whole detector is accelerated by the external force, its structure must be distorted in a way similar to the second case we studied here. It is usually assumed that the detector's structure is not affected by the acceleration, but our result shows that both the wave function and energy of an atom are different from those in an inertial frame. This effect of course depends both on  the magnitude of the acceleration and the size of the atom ($\approx a_0$).

\begin{acknowledgments}
D.C. Dai was supported by the National Science Foundation of China (Grant No. 11433001 and 11447601), National Basic Research Program of China (973 Program 2015CB857001), No.14ZR1423200 from the Office of Science and Technology in Shanghai Municipal Government, the key laboratory grant from the Office of Science and Technology in Shanghai Municipal Government (No. 11DZ2260700) and  the Program of Shanghai Academic/Technology Research Leader under Grant No. 16XD1401600.

\end{acknowledgments}

\appendix*
\section{}

For a particle in a vector field, $B_\alpha$, its classical Lagrangian is
\begin{eqnarray}
S&=&\int -m \sqrt{(1+az)^2-v^2}-B_iv^i-B_t(z) dt\\
B_i&=&0
\end{eqnarray}

Here the coupling constant is included in $B_\alpha$. Vector field $B_\alpha$ keeps a particle rest. In other words it cancels out the gravity or it provides the particle a uniform acceleration with respected to an inertial observer. The equation of motion in z-direction is

\begin{equation}
m\frac{d}{dt}(\frac{v^z}{\sqrt{(1+az)^2-v^2}})=-m\frac{a}{\sqrt{(1+az)^2-v^2}}-\partial_z B_t
\end{equation}

If $\vec{v}=0$, $\frac{dv^z}{dt}=0$. $B_t$ must be

\begin{equation}
B_t=-maz
\end{equation}

Here we choose $B_t=0$ at $z=0$.

\end{document}